\documentclass[aps,prl,reprint,superscriptaddress,showpacs]{revtex4-1}

\usepackage{graphicx}
\usepackage{amssymb,amsmath}
\usepackage{bm}
\usepackage{bbold} 

\begin{document}

\title{Multichannel quantum-defect theory for ion-atom interactions}

\author{Ming Li}
\affiliation{Department of Physics and Astronomy,
    University of Toledo, Mailstop 111,
    Toledo, Ohio 43606,
    USA}
\author{Li You}
\affiliation{State Key Laboratory of Low Dimensional Quantum Physics,
Department of Physics, Tsinghua University, Beijing 100084, China}
\affiliation{Collaborative Innovation Center of Quantum Matter, Beijing, China}
\author{Bo Gao}
\affiliation{Department of Physics and Astronomy,
    University of Toledo, Mailstop 111,
    Toledo, Ohio 43606,
    USA}

\date{February 8, 2014}

\begin{abstract}

We present a quantum theory of ion-atom interaction that is applicable at energies comparable to or smaller than the atomic hyperfine splitting and takes proper account of the effects of identical nuclei. The theory reveals the subtlety and the complexity of cold ion-atom interactions including the change of threshold behavior due to hyperfine splitting and the existence of a large number, and a variety of scattering resonances between hyperfine thresholds. We show how this complexity is described, efficiently and at a quantitative level, using a multichannel quantum-defect theory that we present here for ion-atom interactions. Such an efficient description is a key enabling element to understand few-body and many-body quantum systems involving ions.

\end{abstract}

\pacs{34.10.+x,34.70.+e,34.50.-s,34.50.Cx}

\maketitle

One of the fundamental contributions of cold-atom physics has been its revelation of universal behaviors in quantum many-body \cite{BDZ2008,GPS2008,SU2013} and few-body systems \cite{bra06,gre10,rit11}. Excluding scaling, ultracold atomic systems behave the same with their only differences being characterized by a few parameters such as the scattering length. At a more fundamental level, such universal behaviors have their origin in the universal ultracold two-body interaction as described by the effective range theory \cite{sch47,bla49,bet49,oma61}. Since this theory quickly breaks down at slightly higher energies and at shorter distances, it is natural to ask the question of whether universal behaviors exist beyond the ultracold energy regime and at higher densities, and whether they exist for systems of mixed species of, e.g., atoms, ions, and electrons. These are important questions in physics, the answers to which will determine the degree we can understand the world around us, including phenomena as diverse as reactive processes in atomic collisions, chemical reactions, and high-$T_c$ superconductivity.

The quantum-defect theory (QDT) for atom-atom \cite{mie84a,Burke1998,gao98b,gao01,gao05a,gao08a} and ion-atom interactions \cite{idz09,gao10a,idz11,LG2012,Gao2013c} have served to establish the existence of universal behaviors far beyond the ultracold regime in atomic interactions. 
They imply universal behaviors, over a wide range of temperatures, in processes such as molecular formulation involving ions
\begin{subequations}
\begin{eqnarray}
A^++A+A &\rightarrow& A^++A_2 \;,
\label{eq:react1}\\
	&\rightarrow& A+A_2^+ 
\label{eq:react2} \;.
\end{eqnarray}
\label{eq:react}	
\end{subequations}
Specializing to the H atom, these 3-body reactions constitute mechanisms for H$_2$ formation that are not yet fully understood, and can potentially change the prevailing belief that H$_2$ in interstellar medium are formed mainly on grain surfaces \cite{wat76}. For Rb atom, process~(\ref{eq:react1}) for the formation of Rb$_2$ has been observed in a recent cold-ion experiment \cite{har12}.

While the prospects of few-body theories built upon better understandings of pairwise interactions would seem clear and straightforward \cite{fad60,lov64,alt67,lee07}, their actual implementation and successes depend critically not only on the accuracy of the underlying two-body theories, but also on their efficiency and simplicity, especially in their descriptions of resonances. Two-body theories of such characteristics have not been fully established beyond the ultracold regime for fundamental atomic interactions. They are highly non-trivial for systems with fine or hyperfine structures because of their multichannel characteristics, and especially so for ion-atom interaction because of its rapid energy dependence and generally large number of contributing partial waves \cite{cot00,LG2012,Gao2013c}.
 
In this Letter, we present a multichannel quantum-defect theory (MQDT) for ion-atom interactions and illustrate its application to resonant charge exchange of group I, II, and helium atoms. It is used both to illustrate the complexity of cold ion-atom interactions and to show how such complexity can be described efficiently and quantitatively using MQDT, to establish it as the ion-atom component of future few-body theories involving ions. For the seemingly simple process of resonant charge exchange, earlier numerical theories \cite{cot00,bod08,zha09} have not fully accounted for the effects of hyperfine structure, limiting their range of applicability to approximately 1 K and above. We show that a proper treatment of hyperfine structure leads to qualitatively different behaviors for cold ion-atom interactions including a change of threshold behavior for hyperfine-changing collisions. We further show that the small energy scale associated with ion-atom interactions \cite{gao10a,LG2012,Gao2013c} is such that there exist a large number of resonances: a collection of shape, Feshbach, and diffraction resonances \cite{gao10a,Gao2013c}, even within the small energy interval of a hyperfine splitting ($\sim 0.1$ K). We show how such complexity is fully characterized using MQDT with a small number of parameters such as the atomic polarizability, and the \textit{gerade} and \textit{ungerade} scattering lengths.

Consider the interaction of an atom (group I , II, or He) of nuclear spin $I_1$ in its ground electronic state with an ion of identical nuclei ($I_2=I_1$) also in the ground electronic state. At low energies, the relevant processes, including elastic, $m$-changing, and hyperfine-changing processes, can be described, for group I atoms, by 
\begin{equation}
A(F_{1i},M_{1i})+A^+(F_{2},M_{2i})
\rightarrow A(F_{1j},M_{1j})+A^+(F_{2},M_{2j}) \;,
\label{eq:reaction1}
\end{equation}
and, for group II atoms or He, by
\begin{equation}
A^+(F_{1i},M_{1i})+A(F_{2},M_{2i})
\rightarrow A^+(F_{1j},M_{1j})+A(F_{2},M_{2j}) \;.
\label{eq:reaction2}
\end{equation}
Here $F_1=I_1\pm 1/2$ is the total angular momentum corresponding to the $^2S$ atomic (or ionic) electronic state, and $F_2=I_2=I_1$ is the total angular momentum corresponding to the $^1S$ state. The $M$s are the corresponding magnetic quantum numbers. The sub-indices $i$ and $j$ refer to the internal states before and after the collision. In such an interaction involving only $S$ electrons, the total ``spin'' angular momentum $\bm{F}=\bm{F}_1+\bm{F}_2$, and the relative orbital angular momentum $\bm{l}$, are independently conserved. The scattering amplitudes, for processes of Eqs.~(\ref{eq:reaction1})-(\ref{eq:reaction2}), are given by \cite{gao96}
\begin{align}
 &f\left(\{F_{1i}M_{1i}, F_2M_{2i}\} \bm{k}_i
	\rightarrow \{F_{1j}M_{1j}, F_2M_{2j}\} \bm{k}_j \right) \nonumber\\
= &-\sum_{lmFM} \frac{2\pi i}{(k_ik_j)^{1/2}}
	Y^*_{lm}(\hat{k}_i)Y_{lm}(\hat{k}_j) \nonumber\\
	&\times \langle F_{1j}M_{1j},F_{2}M_{2j}|FM_F\rangle
	\left[S^{Fl}(E)-\mathbb{1}\right]_{ji} \nonumber\\
	&\times\langle FM_F|F_{1i}M_{1i},F_{2}M_{2i}\rangle \;,
\label{eq:scamp}
\end{align}
with the corresponding differential cross section given by \cite{gao96,Gao2013b}
\begin{multline}
\frac{d\sigma}{d\Omega_j}
	\left(\{F_{1i}M_{1i}, F_2M_{2i}\} \bm{k}_i
	\rightarrow \{F_{1j}M_{1j}, F_2M_{2j}\} \bm{k}_j \right) \\
=\frac{k_j}{k_i}\frac{1}{2}\left(\left|f(i\rightarrow j,\bm{k}_j)\right|^2
	+\left|f(i\rightarrow j,-\bm{k}_j)\right|^2\right)\;,
\label{eq:dfxs}
\end{multline}
where $f(i\rightarrow j)$ is a short-hand notation for the amplitude of Eq.~(\ref{eq:scamp}). In Eq.~(\ref{eq:scamp}), $\mathbb{1}$ is the unit matrix and $S^{Fl}$ is the $S$ matrix defined in the $FF$ coupled fragmentation channels that diagonalize the hyperfine interaction \cite{gao96}. 
$\hbar\bm{k}_{i, j}$ are the initial and the final relative momenta in the center-of-mass frame, and $Y_{lm}$ is the spherical harmonics. 
The channel structure is given in Table~\ref{tb:channels}.

\begin{table}
\caption{Channel structure for ion-atom interaction of the type $^2S+^1S$ with identical nuclei of spin $I_2=I_1$. For each partial wave $l$, it consists of a set of two-channel problems for $1/2\le F \le 2I_1-1/2$, and a single-channel problem for $F=2I_1+1/2$. The two $FF$ coupled fragmentation channels, $\{I_1-1/2,I_1\}$ and $\{I_1+1/2,I_1\}$, are separated by the atomic (or ionic) hyperfine splitting $\Delta E^{\textrm{hf}}$. The $JI$ coupling, defined by $\bm{J}=\bm{J}_1+\bm{J}_2$, $\bm{I}=\bm{I}_1+\bm{I}_2$, and $\bm{F}=\bm{J}+\bm{I}$, is used in a frame transformation for the short-range $K^c$ matrix.
    \label{tb:channels}}
\begin{ruledtabular}
\begin{tabular}{ccc}
$F$ & $FF$ coupling $\{F_1,F_2\}$ & $JI$ coupling $\{J,I\}$\\
\hline
$1/2\le F \le 2I_1-1/2$ & $\{I_1-1/2,I_1\}$ & $\{1/2,F-1/2\}$ \\
 & $\{I_1+1/2,I_1\}$ & $\{1/2,F+1/2\}$ \\
\hline $F=2I_1+1/2$ & $\{I_1+1/2,I_1\}$ & $\{1/2,2I_1\}$
\end{tabular}
\end{ruledtabular}
\end{table}

Many different cross sections can be derived. In particular, the total cross sections for elastic (include $m$-changing) collisions, and the hyperfine excitation or de-excitation processes are given by
\begin{align}
\sigma(\{F_{1i},F_{2}\}&\rightarrow\{F_{1j},F_{2}\}) = \frac{\pi}{(2F_{1i}+1)(2F_{2}+1)k_i^2}  \nonumber\\
&\times \sum_{Fl}(2l+1)(2F+1) |S^{Fl}_{ji}-\delta_{ji}|^2 \;.
\label{eq:totalcs}
\end{align}
We will calculate the $S$ matrices both numerically and using MQDT, and compare the resulting cross sections.

In numerical calculations, the coupled-channel (CC) equations in $FF$-coupled basis are set up using the $^2\Sigma_{g,u}^+$ molecular potentials and the atomic hyperfine splitting in a way as outlined in Ref.~\cite{gao96}. The potentials chosen are those constructed in our earlier work \cite{LG2012}. The CC equations are integrated numerically using a hybrid propagator~\cite{man86, ale87}
constructed similarly to the one used in the Hibridon scattering code~\cite{hibridon}.

The MQDT for ion-atom interactions consists of the formulation of Ref.~\cite{gao05a} in combination with the QDT functions for the $-1/R^4$-type potential as detailed in Ref.~\cite{Gao2013c}. 
It takes full advantage of the physics that both the energy dependence \cite{gao98b} and the partial wave dependence \cite{gao01} of the atomic interaction around a threshold are dominated by effects of the long-range potential, which are encapsulated in the universal QDT functions. The short-range contribution is isolated to a short-range $K^c$ matrix that is insensitive to both the energy and the partial wave.
For an $N$-channel problem and at energies where all channels are open, the MQDT gives the physical $K$ matrix, in our case the $K^{Fl}$, as \cite{gao05a}
\begin{equation}
{ K^{Fl}} = -( { Z}^c_{fc}-{ Z}^c_{gc}{ K}^{c} )({ Z}^c_{fs} - { Z}^c_{gs}{ K}^{c})^{-1} \;,
\label{eq:Kphyo}
\end{equation}
where $Z^c_{xy}$s are $N\times N$ diagonal matrices with elements $Z^c_{xy}(\epsilon_{si},l)$ being the $Z^c_{xy}$ functions \cite{Gao2013c} evaluated at scaled energy $\epsilon_{si}=(E-E_i)/s_E$ relative to the respective channel threshold $E_i$. Here $s_E = (\hbar^2/2\mu)(1/\beta_4)^2$ and $\beta_4 = (\mu \alpha_A/\hbar^2)^{1/2}$ are the characteristic energy and the length scales, respectively, associated with the polarization potential, $-\alpha_A/2R^4$, with $\mu$ being the reduced mass and $\alpha_A$ being the static polarizability of the atom. At energies where $N_o$ channels are open, and $N_c=N-N_o$ channels are closed, it gives \cite{gao05a}
\begin{equation} 
K^{Fl} = -( Z^c_{fc}-Z^c_{gc}K^c_{\mathrm{eff}} )(Z^c_{fs} - Z^c_{gs}K^c_{\mathrm{eff}})^{-1} \;,
\label{eq:Kphy}
\end{equation}
where
\begin{equation}
K^c_{\mathrm{eff}} = K^{c}_{oo}+K^{c}_{oc}(\chi^c -K^{c}_{cc})^{-1}K^{c}_{co} \;,
\label{eq:Kceff}
\end{equation}
in which $\chi^c$ is a $N_c\times N_c$ diagonal matrix with elements $\chi^c_l(\epsilon_{si},l)$ \cite{Gao2013c}, and $K^{c}_{oo}$, $K^{c}_{oc}$, $K^{c}_{co}$, and $K^{c}_{cc}$, are submatrices of $K^{c}$ corresponding to open-open, open-closed, closed-open, and closed-closed channels, respectively. Equation~(\ref{eq:Kphy}) is formally the same as Eq.~(\ref{eq:Kphyo}), except that the $K^c$ matrix is replaced by $K^c_{\mathrm{eff}}$ that accounts for the effects of closed channels. From the physical $K$ matrix, the physical $S$ matrix is obtained from ${S^{Fl}} = (\mathbb{1}+i{K^{Fl}})(\mathbb{1}-i{K^{Fl}})^{-1}$~\cite{gao96}. 

This formalism works the same for all group I, II, and He atoms interacting with its corresponding ion. In such an application, the short-range $K^c$ matrix, defined in the $FF$-coupled fragmentation channels, can be obtained from two single-channel $K^c$ matrices, $K^c_{g, u}(\epsilon, l)$, or their corresponding quantum defects, $\mu^c_{g, u} (\epsilon, l)$, through a frame transformation, as detailed in the supplemental material \cite{sup}.
In its simplest implementation, the MQDT describes any one of these systems using three parameters, the atomic polarizability, and the \textit{gerade} and \textit{ungerade} scattering lengths, plus two more atomic properties that are usually well known to great precision: the hyperfine splitting and the atomic mass. When more accurate results are desired over a greater range of energies, the energy and the partial wave dependences of the short-range parameters can be incorporated as needed.  

We illustrate our theory with results for $^{23}\mathrm{Na} + ^{23}\mathrm{Na}^+$, representative of typical behaviors of alkali-metal atoms.
The results for hydrogen, highly relevant in astrophysical applications \cite{Field1958,Glassgold05,Furlanetto2007} including H$_2$ formation mentioned earlier, will be presented elsewhere. The $^{23}$Na has a  nuclear spin of  $I_1 = 3/2$, a hyperfine splitting between $F_1=1$ and 2 states of $\Delta E^{\textrm{hf}}/h \approx 1771.6$ MHz \cite{Arimondo1977} ($\Delta E^{\textrm{hf}}/k_B \approx 0.08502$ K), and a polarizability of $\alpha_A=162.7$ a.u. \cite{Ekstrom1995}.
In the simplest MQDT implementation, $\mu^c_{g,u}$ are taken to be partial wave independent constants $\mu^c_{g}(\epsilon,l)\approx \mu^c_g(0, 0) = 0.42823$ and $\mu^c_{u}(\epsilon,l)\approx \mu^c_u(0, 0) =  0.82922$, corresponding to $s$ wave scattering lengths of $a_{gl=0} = 423.51$ a.u. for the \textit{gerade} state and $a_{ul=0}=-3104.8$ a.u. for the \textit{ungerade} state \cite{LG2012}.
The results, which we call the baseline MQDT results, are illustrated in Figs.~1-3 of the supplemental material \cite{sup}. They show that even the simplest MQDT parametrization gives good descriptions of Na+Na$^+$ up to around 0.4 K, with differences being mainly associated with resonances in high partial waves.

\begin{figure}
\includegraphics[width=\columnwidth]{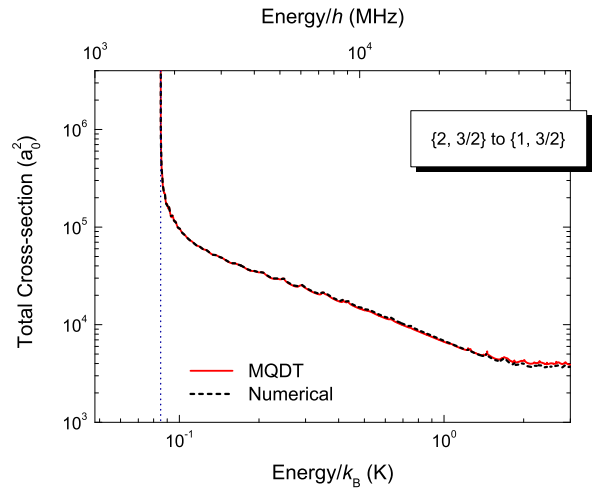}
\caption{ Comparison of the total hyperfine de-excitation cross sections from channel $\{ F_1 = 2, F_2 = 3/2 \}$ to channel $\{ F_1 = 1, F_2 = 3/2 \}$ from MQDT (solid line) and numerical method (dashed line). The vertical dashed line identifies the upper hyperfine threshold located at $E_2/k_B \approx 0.08502$ K, around which the cross section diverges as $(E-E_2)^{-1/2}$. 
\label{fig:10TotalCS}}
\end{figure}

\begin{figure}
\includegraphics[width=\columnwidth]{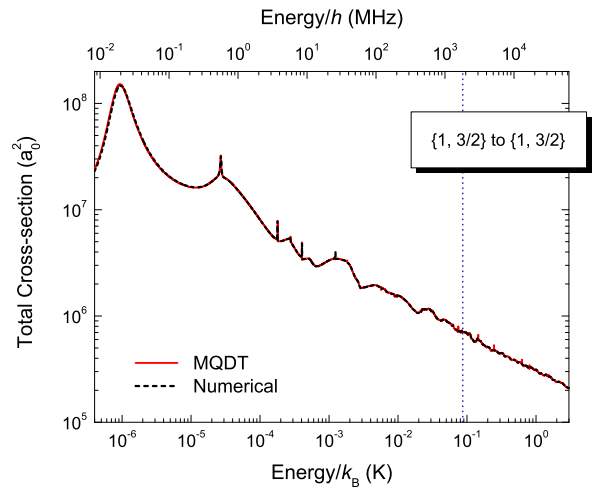}
\caption{ Comparison of the total elastic cross sections in the lower channel $\{ F_1 = 1, F_2 = 3/2 \}$ from MQDT (solid line) and numerical method (dashed line). The vertical dashed line identifies the upper hyperfine threshold at $0.08502$ K. 
\label{fig:00TotalCS}}
\end{figure}

\begin{figure}
\includegraphics[width=\columnwidth]{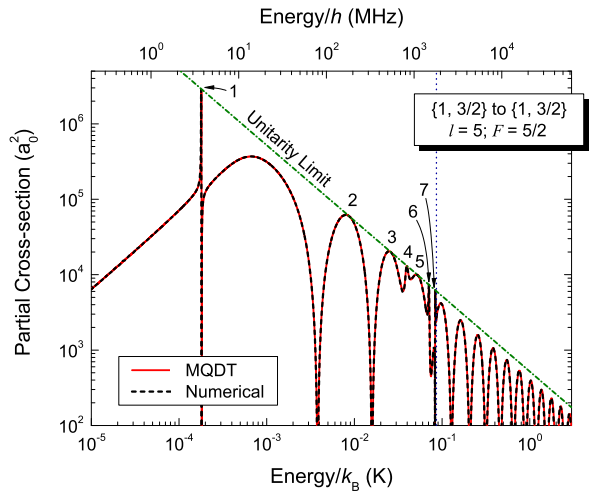}
\caption{Partial wave contribution to the elastic cross section of Fig.~\ref{fig:00TotalCS} from $l = 5$ and $F = 5/2$. There are seven resonances within the hyperfine splitting that are labelled with numbers 1 through 7. Their detailed characteristics are tabulated in Table.~\ref{tb:res}. 
\label{fig:00l5F25}}
\end{figure}

More accurate results over a greater range of energies can be obtained by incorporating the energy dependence, and especially, for the range of energy under consideration, the partial wave dependence of the short-range parameters. These weak dependences are well described by expansions
\begin{equation}
\mu^c_{g, u} (\epsilon, l) \approx \mu^c_{g, u} (0, 0) + b^{\mu}_{g,u}\epsilon+c^{\mu}_{g, u} [l(l + 1)]  \;,
\label{eq:mucel}
\end{equation}
in which the parameters $b^{\mu}_{g,u}$ and $c^{\mu}_{g, u}$ characterize the energy and the partial wave dependences of the quantum defects for the \textit{gerade} and \textit{ungerade} states, respectively. They can be determined easily through single-channel calculations at a few energies and for a few partial waves. For energies up to a few Kelvin, the energy variation of $\mu^c$ is found to be negligible, namely $b^{\mu}_{g,u}\approx 0$, and we find $c^\mu_g = 5.265 \times 10^{-4}$ and $c^\mu_u = 1.030 \times 10^{-3}$.
Figures~\ref{fig:10TotalCS}-\ref{fig:00l5F25} depict the total hyperfine de-excitation, a total elastic, and a sample partial elastic cross sections in which the MQDT results are evaluated with this $l$-dependent $\mu^c$. They show that with the addition of two more parameters (one per channel) that characterize the partial wave dependences of the short-range parameters, MQDT provides quantitatively accurate results that are in full agreement with numerical results and cover the entire energy range of 0 to 3 K in which hyperfine and quantum effects are the most important \cite{sup}.

More specifically, Fig.~\ref{fig:10TotalCS} shows the total cross section for hyperfine de-excitation in which the Na atom is de-excited from its $F_1=2$ hyperfine state to its $F_1=1$ hyperfine state. In earlier studies of resonant charge exchange \cite{cot00,bod08,zha09} using elastic approximation \cite{dal61}, the de-excitation cross section goes to a constant at the threshold \cite{cot00,Glassgold05,Furlanetto2007}. Figure~\ref{fig:10TotalCS} shows the altered threshold behavior with the proper treatment of the hyperfine structure. The de-excitation cross section behaves as $(E-E_2)^{-1/2}$ above the upper threshold, implying a constant rate in the zero temperature limit, as opposed to a zero rate. The hyperfine excitation cross section can be obtained from a detailed balance relation, implied in Eq.~(\ref{eq:totalcs}) and guaranteed by the time-reversal symmetry \cite{lan77}.

Figure~\ref{fig:00TotalCS} depicts the total cross sections for elastic scattering in the lower channel $\{ F_1 = 1, F_2 = 3/2 \}$, in which the atom stays in the lower hyperfine level (but its $M_1$ may change). It shows the complexity of ion-atom interaction as a result of the rapid energy variation induced by the long-range polarization potential. Even within a small energy interval of a hyperfine splitting ($\sim 0.085$ K), the small energy scale associated with the long-range potential, $s_E\approx 2.21$ $\mu$K, is such that there are many contributing partial waves (more than $\sqrt{2}(\epsilon/s_E)^{1/4}\sim 20$ \cite{LG2012} at the upper threshold). Each partial wave contribution contains  a variety of resonances, which, in the energy region below the second threshold, include Feshbach resonances in addition to shape and diffraction resonances~\cite{gao10a,Gao2013c}.

\begin{table}
\caption{Positions, widths, and classifications of the 7 resonances labeled in Fig.~\ref{fig:00l5F25}. 
\label{tb:res}}
\begin{ruledtabular}
\begin{tabular}{cccc}
Resonance & Energy$/ k_{\rm B}$ (K) & Width$/h$ (MHz) & Type \\
\hline
1 & $1.801\times 10^{-4}$ & $1.562\times 10^{-2}$ & Shape \\
2 & $8.795\times 10^{-3}$ & $-1.573\times 10^{2}$ & Diffraction \\
3 & $2.600\times 10^{-2}$ & $-3.235\times 10^{2}$ & Diffraction \\
4 & $3.970\times 10^{-2}$ & $1.030\times 10^{2}$ & Feshbach \\
5 & $5.271\times 10^{-2}$ & $-5.669\times 10^{2}$ & Diffraction \\
6 & $7.085\times 10^{-2}$ & $3.946\times 10^{1}$ & Feshbach \\
7 & $8.370\times 10^{-2}$ & $1.647\times 10^{1}$ & Feshbach
\end{tabular}
\end{ruledtabular}
\end{table}

Figure~\ref{fig:00l5F25} depicts the partial wave contribution from $l=5$ and $F=5$, for which we illustrate MQDT analysis of the resonances. There are seven resonances between the two hyperfine thresholds. The resonance positions, which can be defined in this region as the energies at which the cross section reaches its unitarity limit, can be found as the solutions of
\begin{equation}
\widetilde{\chi}^c_l(\epsilon_s) - K^c_{\mathrm{eff}} = 0 \;,
\label{eq:resOneCh}
\end{equation}
where $\widetilde{\chi}^c_l(\epsilon_s) = Z^c_{fs}/Z^c_{gs}$ \cite{Gao2013c}. The widths of the resonances, $\Gamma_l$, more specically the scaled widths, $\Gamma_{sl}\equiv \Gamma_l/s_E$, can be shown to be given by
\begin{equation}
\Gamma_{sl} = -\frac{2}{ \left[ Z^c_{gs} (\epsilon_{sl}, l) \right]^2 \left[ \left.\frac{\mathrm{d}\widetilde{\chi}^c_l}{\mathrm{d}\epsilon_s}\right|_{\epsilon_{s1l}} + \left. \frac{K^{c}_{oc}K^{c}_{co}}{(\chi^c_l - K^{c}_{cc})^2}\frac{\mathrm{d}\chi^c_l}{\mathrm{d}\epsilon_s} \right|_{\epsilon_{s2l}} \right]  } \;,
\label{eq:sWidth}
\end{equation}
where $\epsilon_{s1l}$ and $\epsilon_{s2l}$ are the scaled resonance positions relative to the lower and the upper thresholds, respectively. Equations~(\ref{eq:resOneCh}) and (\ref{eq:sWidth}) are part of the multichannel generalizations of the concept of resonance spectrum and the corresponding width function \cite{Gao2013c}. They describe the pole structures of the physical $K$ matrix that has to be understood efficiently in applications beyond two-body physics \cite{lov64,alt67,lee07}. Table~\ref{tb:res} gives their characterizations of all resonances labeled in Fig.~\ref{fig:00l5F25}. Resonances of negative widths are diffraction resonances corresponding to reductions of density-of-states \cite{gao10a,Gao2013c}. Resonances of positive widths can be either shape or Feshbach resonances. They can be distinguished by comparing their locations with those of the ``bare'' shape resonances associated with the lower open channel and the ``bare'' Feshbach resonances associated with the upper closed channel. We note that while the resonances in a single partial wave tend to be smeared in the total cross section (Fig.~\ref{fig:00TotalCS}) due to the summation over a large number of contributing partial waves, they are in principle  observable in photodissociation of a molecular ion \cite{hec02} from an excited rovibrational state.

In conclusion, we have presented a MQDT for ion-atom interactions and illustrated its application to resonant charge exchange. The theory provides a systematic and basically an analytic description of ion-atom systems in its most complex energy regime where quantum effects are important. Other than well-known atomic properties such as the atomic mass, hyperfine splitting, and the atomic polarizability, different group I, II, and He atoms differ primarily only in two parameters such as the \textit{gerade} and \textit{ungerade} scattering lengths or quantum defects, and secondarily (when interest is over a greater range of energies) in two more parameters, $c^{\mu}_{g, u}$, that characterize their partial wave dependences. Theoretically, MQDT gives a complete understanding and characterization of threshold behaviors and complex resonance structures, and helps to overcome the sensitive dependence of cold atomic interactions on short-range potentials \cite{LG2012}.
Computationally, MQDT is much more efficient than numerical calculations even when the QDT functions are calculated on the fly. Since the QDT functions are universal mathematical functions that are the same for all applications, and can be computed to arbitrary precision with efficient algorithms \cite{Gao2013c}, their computation can be further accelerated to be as efficient as most other mathematical special functions.
We believe that the systematic and efficient understanding of ion-atom interactions that our theory provides, especially in the cold temperature regime where quantum effects are important, will be the key to systematic understanding of quantum few-body systems, chemical reactions, and many-body systems involving ions.

We thank Dr.~Meng Khoon Tey for helpful discussions. The work is supported at Toledo by NSF (PHY-1306407), and at Tsinghua by NSFC (No.~91121005, No.~11374176, and No.~11328404) and by MOST 2013CB922004 of the National Key Basic Research Program of China.

\bibliography{bgao,qdt,ionAtom,twobody,fewbody,manybody,atom,numerical,astrochem}

\begin{thebibliography}{44}%
\makeatletter
\providecommand \@ifxundefined [1]{%
 \@ifx{#1\undefined}
}%
\providecommand \@ifnum [1]{%
 \ifnum #1\expandafter \@firstoftwo
 \else \expandafter \@secondoftwo
 \fi
}%
\providecommand \@ifx [1]{%
 \ifx #1\expandafter \@firstoftwo
 \else \expandafter \@secondoftwo
 \fi
}%
\providecommand \natexlab [1]{#1}%
\providecommand \enquote  [1]{``#1''}%
\providecommand \bibnamefont  [1]{#1}%
\providecommand \bibfnamefont [1]{#1}%
\providecommand \citenamefont [1]{#1}%
\providecommand \href@noop [0]{\@secondoftwo}%
\providecommand \href [0]{\begingroup \@sanitize@url \@href}%
\providecommand \@href[1]{\@@startlink{#1}\@@href}%
\providecommand \@@href[1]{\endgroup#1\@@endlink}%
\providecommand \@sanitize@url [0]{\catcode `\\12\catcode `\$12\catcode
  `\&12\catcode `\#12\catcode `\^12\catcode `\_12\catcode `\%12\relax}%
\providecommand \@@startlink[1]{}%
\providecommand \@@endlink[0]{}%
\providecommand \url  [0]{\begingroup\@sanitize@url \@url }%
\providecommand \@url [1]{\endgroup\@href {#1}{\urlprefix }}%
\providecommand \urlprefix  [0]{URL }%
\providecommand \Eprint [0]{\href }%
\providecommand \doibase [0]{http://dx.doi.org/}%
\providecommand \selectlanguage [0]{\@gobble}%
\providecommand \bibinfo  [0]{\@secondoftwo}%
\providecommand \bibfield  [0]{\@secondoftwo}%
\providecommand \translation [1]{[#1]}%
\providecommand \BibitemOpen [0]{}%
\providecommand \bibitemStop [0]{}%
\providecommand \bibitemNoStop [0]{.\EOS\space}%
\providecommand \EOS [0]{\spacefactor3000\relax}%
\providecommand \BibitemShut  [1]{\csname bibitem#1\endcsname}%
\let\auto@bib@innerbib\@empty
\bibitem [{\citenamefont {Bloch}\ \emph {et~al.}(2008)\citenamefont {Bloch},
  \citenamefont {Dalibard},\ and\ \citenamefont {Zwerger}}]{BDZ2008}%
  \BibitemOpen
  \bibfield  {author} {\bibinfo {author} {\bibfnamefont {I.}~\bibnamefont
  {Bloch}}, \bibinfo {author} {\bibfnamefont {J.}~\bibnamefont {Dalibard}}, \
  and\ \bibinfo {author} {\bibfnamefont {W.}~\bibnamefont {Zwerger}},\ }\href
  {\doibase 10.1103/RevModPhys.80.885} {\bibfield  {journal} {\bibinfo
  {journal} {Rev. Mod. Phys.}\ }\textbf {\bibinfo {volume} {80}},\ \bibinfo
  {pages} {885} (\bibinfo {year} {2008})}\BibitemShut {NoStop}%
\bibitem [{\citenamefont {Giorgini}\ \emph {et~al.}(2008)\citenamefont
  {Giorgini}, \citenamefont {Pitaevskii},\ and\ \citenamefont
  {Stringari}}]{GPS2008}%
  \BibitemOpen
  \bibfield  {author} {\bibinfo {author} {\bibfnamefont {S.}~\bibnamefont
  {Giorgini}}, \bibinfo {author} {\bibfnamefont {L.~P.}\ \bibnamefont
  {Pitaevskii}}, \ and\ \bibinfo {author} {\bibfnamefont {S.}~\bibnamefont
  {Stringari}},\ }\href {\doibase 10.1103/RevModPhys.80.1215} {\bibfield
  {journal} {\bibinfo  {journal} {Rev. Mod. Phys.}\ }\textbf {\bibinfo {volume}
  {80}},\ \bibinfo {pages} {1215} (\bibinfo {year} {2008})}\BibitemShut
  {NoStop}%
\bibitem [{\citenamefont {Stamper-Kurn}\ and\ \citenamefont
  {Ueda}(2013)}]{SU2013}%
  \BibitemOpen
  \bibfield  {author} {\bibinfo {author} {\bibfnamefont {D.~M.}\ \bibnamefont
  {Stamper-Kurn}}\ and\ \bibinfo {author} {\bibfnamefont {M.}~\bibnamefont
  {Ueda}},\ }\href {\doibase 10.1103/RevModPhys.85.1191} {\bibfield  {journal}
  {\bibinfo  {journal} {Rev. Mod. Phys.}\ }\textbf {\bibinfo {volume} {85}},\
  \bibinfo {pages} {1191} (\bibinfo {year} {2013})}\BibitemShut {NoStop}%
\bibitem [{\citenamefont {Braaten}\ and\ \citenamefont {Hammer}(2006)}]{bra06}%
  \BibitemOpen
  \bibfield  {author} {\bibinfo {author} {\bibfnamefont {E.}~\bibnamefont
  {Braaten}}\ and\ \bibinfo {author} {\bibfnamefont {H.-W.}\ \bibnamefont
  {Hammer}},\ }\href {\doibase DOI: 10.1016/j.physrep.2006.03.001} {\bibfield
  {journal} {\bibinfo  {journal} {Physics Reports}\ }\textbf {\bibinfo {volume}
  {428}},\ \bibinfo {pages} {259 } (\bibinfo {year} {2006})}\BibitemShut
  {NoStop}%
\bibitem [{\citenamefont {Greene}(2010)}]{gre10}%
  \BibitemOpen
  \bibfield  {author} {\bibinfo {author} {\bibfnamefont {C.~H.}\ \bibnamefont
  {Greene}},\ }\href@noop {} {\bibfield  {journal} {\bibinfo  {journal}
  {Physics Today}\ }\textbf {\bibinfo {volume} {63}},\ \bibinfo {pages} {40}
  (\bibinfo {year} {2010})}\BibitemShut {NoStop}%
\bibitem [{\citenamefont {Rittenhouse}\ \emph {et~al.}(2011)\citenamefont
  {Rittenhouse}, \citenamefont {von Stecher}, \citenamefont {D'Incao},
  \citenamefont {Mehta},\ and\ \citenamefont {Greene}}]{rit11}%
  \BibitemOpen
  \bibfield  {author} {\bibinfo {author} {\bibfnamefont {S.~T.}\ \bibnamefont
  {Rittenhouse}}, \bibinfo {author} {\bibfnamefont {J.}~\bibnamefont {von
  Stecher}}, \bibinfo {author} {\bibfnamefont {J.~P.}\ \bibnamefont {D'Incao}},
  \bibinfo {author} {\bibfnamefont {N.~P.}\ \bibnamefont {Mehta}}, \ and\
  \bibinfo {author} {\bibfnamefont {C.~H.}\ \bibnamefont {Greene}},\ }\href
  {http://stacks.iop.org/0953-4075/44/i=17/a=172001} {\bibfield  {journal}
  {\bibinfo  {journal} {Journal of Physics B: Atomic, Molecular and Optical
  Physics}\ }\textbf {\bibinfo {volume} {44}},\ \bibinfo {pages} {172001}
  (\bibinfo {year} {2011})}\BibitemShut {NoStop}%
\bibitem [{\citenamefont {Schwinger}(1947)}]{sch47}%
  \BibitemOpen
  \bibfield  {author} {\bibinfo {author} {\bibfnamefont {J.}~\bibnamefont
  {Schwinger}},\ }\href {\doibase 10.1103/PhysRev.72.738} {\bibfield  {journal}
  {\bibinfo  {journal} {Phys. Rev.}\ }\textbf {\bibinfo {volume} {72}},\
  \bibinfo {pages} {738} (\bibinfo {year} {1947})}\BibitemShut {NoStop}%
\bibitem [{\citenamefont {Blatt}\ and\ \citenamefont {Jackson}(1949)}]{bla49}%
  \BibitemOpen
  \bibfield  {author} {\bibinfo {author} {\bibfnamefont {J.~M.}\ \bibnamefont
  {Blatt}}\ and\ \bibinfo {author} {\bibfnamefont {D.~J.}\ \bibnamefont
  {Jackson}},\ }\href@noop {} {\bibfield  {journal} {\bibinfo  {journal} {Phys.
  Rev.}\ }\textbf {\bibinfo {volume} {76}},\ \bibinfo {pages} {18} (\bibinfo
  {year} {1949})}\BibitemShut {NoStop}%
\bibitem [{\citenamefont {Bethe}(1949)}]{bet49}%
  \BibitemOpen
  \bibfield  {author} {\bibinfo {author} {\bibfnamefont {H.~A.}\ \bibnamefont
  {Bethe}},\ }\href {\doibase 10.1103/PhysRev.76.38} {\bibfield  {journal}
  {\bibinfo  {journal} {Phys. Rev.}\ }\textbf {\bibinfo {volume} {76}},\
  \bibinfo {pages} {38} (\bibinfo {year} {1949})}\BibitemShut {NoStop}%
\bibitem [{\citenamefont {O'Malley}\ \emph {et~al.}(1961)\citenamefont
  {O'Malley}, \citenamefont {Spruch},\ and\ \citenamefont {Rosenberg}}]{oma61}%
  \BibitemOpen
  \bibfield  {author} {\bibinfo {author} {\bibfnamefont {T.~F.}\ \bibnamefont
  {O'Malley}}, \bibinfo {author} {\bibfnamefont {L.}~\bibnamefont {Spruch}}, \
  and\ \bibinfo {author} {\bibfnamefont {L.}~\bibnamefont {Rosenberg}},\ }\href
  {\doibase 10.1063/1.1703735} {\bibfield  {journal} {\bibinfo  {journal} {J.
  Math. Phys.}\ }\textbf {\bibinfo {volume} {2}},\ \bibinfo {pages} {491}
  (\bibinfo {year} {1961})}\BibitemShut {NoStop}%
\bibitem [{\citenamefont {Mies}(1984)}]{mie84a}%
  \BibitemOpen
  \bibfield  {author} {\bibinfo {author} {\bibfnamefont {F.~H.}\ \bibnamefont
  {Mies}},\ }\href {\doibase 10.1063/1.447000} {\bibfield  {journal} {\bibinfo
  {journal} {The Journal of Chemical Physics}\ }\textbf {\bibinfo {volume}
  {80}},\ \bibinfo {pages} {2514} (\bibinfo {year} {1984})}\BibitemShut
  {NoStop}%
\bibitem [{\citenamefont {Burke}\ \emph {et~al.}(1998)\citenamefont {Burke},
  \citenamefont {Greene},\ and\ \citenamefont {Bohn}}]{Burke1998}%
  \BibitemOpen
  \bibfield  {author} {\bibinfo {author} {\bibfnamefont {J.~P.}\ \bibnamefont
  {Burke}}, \bibinfo {author} {\bibfnamefont {C.~H.}\ \bibnamefont {Greene}}, \
  and\ \bibinfo {author} {\bibfnamefont {J.~L.}\ \bibnamefont {Bohn}},\ }\href
  {\doibase 10.1103/PhysRevLett.81.3355} {\bibfield  {journal} {\bibinfo
  {journal} {Phys. Rev. Lett.}\ }\textbf {\bibinfo {volume} {81}},\ \bibinfo
  {pages} {3355} (\bibinfo {year} {1998})}\BibitemShut {NoStop}%
\bibitem [{\citenamefont {Gao}(1998)}]{gao98b}%
  \BibitemOpen
  \bibfield  {author} {\bibinfo {author} {\bibfnamefont {B.}~\bibnamefont
  {Gao}},\ }\href@noop {} {\bibfield  {journal} {\bibinfo  {journal} {Phys.
  Rev. A}\ }\textbf {\bibinfo {volume} {58}},\ \bibinfo {pages} {4222}
  (\bibinfo {year} {1998})}\BibitemShut {NoStop}%
\bibitem [{\citenamefont {Gao}(2001)}]{gao01}%
  \BibitemOpen
  \bibfield  {author} {\bibinfo {author} {\bibfnamefont {B.}~\bibnamefont
  {Gao}},\ }\href@noop {} {\bibfield  {journal} {\bibinfo  {journal} {Phys.
  Rev. A}\ }\textbf {\bibinfo {volume} {64}},\ \bibinfo {pages} {010701(R)}
  (\bibinfo {year} {2001})}\BibitemShut {NoStop}%
\bibitem [{\citenamefont {Gao}\ \emph {et~al.}(2005)\citenamefont {Gao},
  \citenamefont {Tiesinga}, \citenamefont {Williams},\ and\ \citenamefont
  {Julienne}}]{gao05a}%
  \BibitemOpen
  \bibfield  {author} {\bibinfo {author} {\bibfnamefont {B.}~\bibnamefont
  {Gao}}, \bibinfo {author} {\bibfnamefont {E.}~\bibnamefont {Tiesinga}},
  \bibinfo {author} {\bibfnamefont {C.~J.}\ \bibnamefont {Williams}}, \ and\
  \bibinfo {author} {\bibfnamefont {P.~S.}\ \bibnamefont {Julienne}},\
  }\href@noop {} {\bibfield  {journal} {\bibinfo  {journal} {Phys. Rev. A}\
  }\textbf {\bibinfo {volume} {72}},\ \bibinfo {pages} {042719} (\bibinfo
  {year} {2005})}\BibitemShut {NoStop}%
\bibitem [{\citenamefont {Gao}(2008)}]{gao08a}%
  \BibitemOpen
  \bibfield  {author} {\bibinfo {author} {\bibfnamefont {B.}~\bibnamefont
  {Gao}},\ }\href@noop {} {\bibfield  {journal} {\bibinfo  {journal} {Phys.
  Rev. A}\ }\textbf {\bibinfo {volume} {78}},\ \bibinfo {pages} {012702}
  (\bibinfo {year} {2008})}\BibitemShut {NoStop}%
\bibitem [{\citenamefont {Idziaszek}\ \emph {et~al.}(2009)\citenamefont
  {Idziaszek}, \citenamefont {Calarco}, \citenamefont {Julienne},\ and\
  \citenamefont {Simoni}}]{idz09}%
  \BibitemOpen
  \bibfield  {author} {\bibinfo {author} {\bibfnamefont {Z.}~\bibnamefont
  {Idziaszek}}, \bibinfo {author} {\bibfnamefont {T.}~\bibnamefont {Calarco}},
  \bibinfo {author} {\bibfnamefont {P.~S.}\ \bibnamefont {Julienne}}, \ and\
  \bibinfo {author} {\bibfnamefont {A.}~\bibnamefont {Simoni}},\ }\href
  {\doibase 10.1103/PhysRevA.79.010702} {\bibfield  {journal} {\bibinfo
  {journal} {Phys. Rev. A}\ }\textbf {\bibinfo {volume} {79}},\ \bibinfo {eid}
  {010702(R)} (\bibinfo {year} {2009})}\BibitemShut {NoStop}%
\bibitem [{\citenamefont {Gao}(2010)}]{gao10a}%
  \BibitemOpen
  \bibfield  {author} {\bibinfo {author} {\bibfnamefont {B.}~\bibnamefont
  {Gao}},\ }\href {\doibase 10.1103/PhysRevLett.104.213201} {\bibfield
  {journal} {\bibinfo  {journal} {Phys. Rev. Lett.}\ }\textbf {\bibinfo
  {volume} {104}},\ \bibinfo {pages} {213201} (\bibinfo {year}
  {2010})}\BibitemShut {NoStop}%
\bibitem [{\citenamefont {Idziaszek}\ \emph {et~al.}(2011)\citenamefont
  {Idziaszek}, \citenamefont {Simoni}, \citenamefont {Calarco},\ and\
  \citenamefont {Julienne}}]{idz11}%
  \BibitemOpen
  \bibfield  {author} {\bibinfo {author} {\bibfnamefont {Z.}~\bibnamefont
  {Idziaszek}}, \bibinfo {author} {\bibfnamefont {A.}~\bibnamefont {Simoni}},
  \bibinfo {author} {\bibfnamefont {T.}~\bibnamefont {Calarco}}, \ and\
  \bibinfo {author} {\bibfnamefont {P.~S.}\ \bibnamefont {Julienne}},\ }\href
  {http://stacks.iop.org/1367-2630/13/i=8/a=083005} {\bibfield  {journal}
  {\bibinfo  {journal} {New Journal of Physics}\ }\textbf {\bibinfo {volume}
  {13}},\ \bibinfo {pages} {083005} (\bibinfo {year} {2011})}\BibitemShut
  {NoStop}%
\bibitem [{\citenamefont {Li}\ and\ \citenamefont {Gao}(2012)}]{LG2012}%
  \BibitemOpen
  \bibfield  {author} {\bibinfo {author} {\bibfnamefont {M.}~\bibnamefont
  {Li}}\ and\ \bibinfo {author} {\bibfnamefont {B.}~\bibnamefont {Gao}},\
  }\href {\doibase 10.1103/PhysRevA.86.012707} {\bibfield  {journal} {\bibinfo
  {journal} {Phys. Rev. A}\ }\textbf {\bibinfo {volume} {86}},\ \bibinfo
  {pages} {012707} (\bibinfo {year} {2012})}\BibitemShut {NoStop}%
\bibitem [{\citenamefont {Gao}(2013{\natexlab{a}})}]{Gao2013c}%
  \BibitemOpen
  \bibfield  {author} {\bibinfo {author} {\bibfnamefont {B.}~\bibnamefont
  {Gao}},\ }\href {\doibase 10.1103/PhysRevA.88.022701} {\bibfield  {journal}
  {\bibinfo  {journal} {Phys. Rev. A}\ }\textbf {\bibinfo {volume} {88}},\
  \bibinfo {pages} {022701} (\bibinfo {year} {2013}{\natexlab{a}})}\BibitemShut
  {NoStop}%
\bibitem [{\citenamefont {Watson}(1976)}]{wat76}%
  \BibitemOpen
  \bibfield  {author} {\bibinfo {author} {\bibfnamefont {W.~D.}\ \bibnamefont
  {Watson}},\ }\href {\doibase 10.1103/RevModPhys.48.513} {\bibfield  {journal}
  {\bibinfo  {journal} {Rev. Mod. Phys.}\ }\textbf {\bibinfo {volume} {48}},\
  \bibinfo {pages} {513} (\bibinfo {year} {1976})}\BibitemShut {NoStop}%
\bibitem [{\citenamefont {H\"arter}\ \emph {et~al.}(2012)\citenamefont
  {H\"arter}, \citenamefont {Kr\"ukow}, \citenamefont {Brunner}, \citenamefont
  {Schnitzler}, \citenamefont {Schmid},\ and\ \citenamefont
  {Denschlag}}]{har12}%
  \BibitemOpen
  \bibfield  {author} {\bibinfo {author} {\bibfnamefont {A.}~\bibnamefont
  {H\"arter}}, \bibinfo {author} {\bibfnamefont {A.}~\bibnamefont {Kr\"ukow}},
  \bibinfo {author} {\bibfnamefont {A.}~\bibnamefont {Brunner}}, \bibinfo
  {author} {\bibfnamefont {W.}~\bibnamefont {Schnitzler}}, \bibinfo {author}
  {\bibfnamefont {S.}~\bibnamefont {Schmid}}, \ and\ \bibinfo {author}
  {\bibfnamefont {J.~H.}\ \bibnamefont {Denschlag}},\ }\href {\doibase
  10.1103/PhysRevLett.109.123201} {\bibfield  {journal} {\bibinfo  {journal}
  {Phys. Rev. Lett.}\ }\textbf {\bibinfo {volume} {109}},\ \bibinfo {pages}
  {123201} (\bibinfo {year} {2012})}\BibitemShut {NoStop}%
\bibitem [{\citenamefont {Faddeev}(1960)}]{fad60}%
  \BibitemOpen
  \bibfield  {author} {\bibinfo {author} {\bibfnamefont {L.}~\bibnamefont
  {Faddeev}},\ }\href@noop {} {\bibfield  {journal} {\bibinfo  {journal} {Zh.
  Eksp. Teor. Fiz.}\ }\textbf {\bibinfo {volume} {39}},\ \bibinfo {pages}
  {1459} (\bibinfo {year} {1960})},\ \bibinfo {note} {[Sov. Phys. JETP
  \textbf{12}, 1014 (1961)]}\BibitemShut {NoStop}%
\bibitem [{\citenamefont {Lovelace}(1964)}]{lov64}%
  \BibitemOpen
  \bibfield  {author} {\bibinfo {author} {\bibfnamefont {C.}~\bibnamefont
  {Lovelace}},\ }\href {\doibase 10.1103/PhysRev.135.B1225} {\bibfield
  {journal} {\bibinfo  {journal} {Phys. Rev.}\ }\textbf {\bibinfo {volume}
  {135}},\ \bibinfo {pages} {B1225} (\bibinfo {year} {1964})}\BibitemShut
  {NoStop}%
\bibitem [{\citenamefont {Alt}\ \emph {et~al.}(1967)\citenamefont {Alt},
  \citenamefont {Grassberger},\ and\ \citenamefont {Sandhas}}]{alt67}%
  \BibitemOpen
  \bibfield  {author} {\bibinfo {author} {\bibfnamefont {E.}~\bibnamefont
  {Alt}}, \bibinfo {author} {\bibfnamefont {P.}~\bibnamefont {Grassberger}}, \
  and\ \bibinfo {author} {\bibfnamefont {W.}~\bibnamefont {Sandhas}},\ }\href
  {\doibase 10.1016/0550-3213(67)90016-8} {\bibfield  {journal} {\bibinfo
  {journal} {Nuclear Physics B}\ }\textbf {\bibinfo {volume} {2}},\ \bibinfo
  {pages} {167 } (\bibinfo {year} {1967})}\BibitemShut {NoStop}%
\bibitem [{\citenamefont {Lee}\ \emph {et~al.}(2007)\citenamefont {Lee},
  \citenamefont {Kohler},\ and\ \citenamefont {Julienne}}]{lee07}%
  \BibitemOpen
  \bibfield  {author} {\bibinfo {author} {\bibfnamefont {M.~D.}\ \bibnamefont
  {Lee}}, \bibinfo {author} {\bibfnamefont {T.}~\bibnamefont {Kohler}}, \ and\
  \bibinfo {author} {\bibfnamefont {P.~S.}\ \bibnamefont {Julienne}},\ }\href
  {\doibase 10.1103/PhysRevA.76.012720} {\bibfield  {journal} {\bibinfo
  {journal} {Phys. Rev. A}\ }\textbf {\bibinfo {volume} {76}},\ \bibinfo {eid}
  {012720} (\bibinfo {year} {2007})}\BibitemShut {NoStop}%
\bibitem [{\citenamefont {C\^ot\'e}\ and\ \citenamefont
  {Dalgarno}(2000)}]{cot00}%
  \BibitemOpen
  \bibfield  {author} {\bibinfo {author} {\bibfnamefont {R.}~\bibnamefont
  {C\^ot\'e}}\ and\ \bibinfo {author} {\bibfnamefont {A.}~\bibnamefont
  {Dalgarno}},\ }\href {\doibase 10.1103/PhysRevA.62.012709} {\bibfield
  {journal} {\bibinfo  {journal} {Phys. Rev. A}\ }\textbf {\bibinfo {volume}
  {62}},\ \bibinfo {pages} {012709} (\bibinfo {year} {2000})}\BibitemShut
  {NoStop}%
\bibitem [{\citenamefont {Bodo}\ \emph {et~al.}(2008)\citenamefont {Bodo},
  \citenamefont {Zhang},\ and\ \citenamefont {Dalgarno}}]{bod08}%
  \BibitemOpen
  \bibfield  {author} {\bibinfo {author} {\bibfnamefont {E.}~\bibnamefont
  {Bodo}}, \bibinfo {author} {\bibfnamefont {P.}~\bibnamefont {Zhang}}, \ and\
  \bibinfo {author} {\bibfnamefont {A.}~\bibnamefont {Dalgarno}},\ }\href
  {http://stacks.iop.org/1367-2630/10/i=3/a=033024} {\bibfield  {journal}
  {\bibinfo  {journal} {New Journal of Physics}\ }\textbf {\bibinfo {volume}
  {10}},\ \bibinfo {pages} {033024} (\bibinfo {year} {2008})}\BibitemShut
  {NoStop}%
\bibitem [{\citenamefont {Zhang}\ \emph {et~al.}(2009)\citenamefont {Zhang},
  \citenamefont {Dalgarno},\ and\ \citenamefont {C\^ot\'e}}]{zha09}%
  \BibitemOpen
  \bibfield  {author} {\bibinfo {author} {\bibfnamefont {P.}~\bibnamefont
  {Zhang}}, \bibinfo {author} {\bibfnamefont {A.}~\bibnamefont {Dalgarno}}, \
  and\ \bibinfo {author} {\bibfnamefont {R.}~\bibnamefont {C\^ot\'e}},\ }\href
  {\doibase 10.1103/PhysRevA.80.030703} {\bibfield  {journal} {\bibinfo
  {journal} {Phys. Rev. A}\ }\textbf {\bibinfo {volume} {80}},\ \bibinfo
  {pages} {030703(R)} (\bibinfo {year} {2009})}\BibitemShut {NoStop}%
\bibitem [{\citenamefont {Gao}(1996)}]{gao96}%
  \BibitemOpen
  \bibfield  {author} {\bibinfo {author} {\bibfnamefont {B.}~\bibnamefont
  {Gao}},\ }\href@noop {} {\bibfield  {journal} {\bibinfo  {journal} {Phys.
  Rev. A}\ }\textbf {\bibinfo {volume} {54}},\ \bibinfo {pages} {2022}
  (\bibinfo {year} {1996})}\BibitemShut {NoStop}%
\bibitem [{\citenamefont {Gao}(2013{\natexlab{b}})}]{Gao2013b}%
  \BibitemOpen
  \bibfield  {author} {\bibinfo {author} {\bibfnamefont {B.}~\bibnamefont
  {Gao}},\ }\href {\doibase 10.1103/PhysRevA.87.059903} {\bibfield  {journal}
  {\bibinfo  {journal} {Phys. Rev. A}\ }\textbf {\bibinfo {volume} {87}},\
  \bibinfo {pages} {059903} (\bibinfo {year} {2013}{\natexlab{b}})}\BibitemShut
  {NoStop}%
\bibitem [{\citenamefont {Manolopoulos}(1986)}]{man86}%
  \BibitemOpen
  \bibfield  {author} {\bibinfo {author} {\bibfnamefont {D.~E.}\ \bibnamefont
  {Manolopoulos}},\ }\href {\doibase 10.1063/1.451472} {\bibfield  {journal}
  {\bibinfo  {journal} {The Journal of Chemical Physics}\ }\textbf {\bibinfo
  {volume} {85}},\ \bibinfo {pages} {6425} (\bibinfo {year}
  {1986})}\BibitemShut {NoStop}%
\bibitem [{\citenamefont {Alexander}\ and\ \citenamefont
  {Manolopoulos}(1987)}]{ale87}%
  \BibitemOpen
  \bibfield  {author} {\bibinfo {author} {\bibfnamefont {M.~H.}\ \bibnamefont
  {Alexander}}\ and\ \bibinfo {author} {\bibfnamefont {D.~E.}\ \bibnamefont
  {Manolopoulos}},\ }\href {\doibase 10.1063/1.452154} {\bibfield  {journal}
  {\bibinfo  {journal} {The Journal of Chemical Physics}\ }\textbf {\bibinfo
  {volume} {86}},\ \bibinfo {pages} {2044} (\bibinfo {year}
  {1987})}\BibitemShut {NoStop}%
\bibitem [{hib()}]{hibridon}%
  \BibitemOpen
  \href@noop {} {}\bibinfo {note} {HIBRIDON$^\copyright$ is a package of
  programs for the time-independent quantum treatment of inelastic collisions
  and photodissociation written by M. H. Alexander, D. E. Manolopoulos, H.-J.
  Werner, and B. Follmeg, with contributions by P. F. Vohralik, D. Lemoine, G.
  Corey, R. Gordon, B. Johnson, T. Orlikowski, A. Berning, A. Degli-Esposti, C.
  Rist, P. Dagdigian, B. Pouilly, G. van der Sanden, M. Yang, F. de Weerd, S.
  Gregurick, and J. Klos.}\BibitemShut {Stop}%
\bibitem [{sup()}]{sup}%
  \BibitemOpen
  \href@noop {} {}\bibinfo {note} {See EPAPS Document No.~x}\BibitemShut
  {NoStop}%
\bibitem [{\citenamefont {Field}(1958)}]{Field1958}%
  \BibitemOpen
  \bibfield  {author} {\bibinfo {author} {\bibfnamefont {G.}~\bibnamefont
  {Field}},\ }\href {\doibase 10.1109/JRPROC.1958.286741} {\bibfield  {journal}
  {\bibinfo  {journal} {Proceedings of the IRE}\ }\textbf {\bibinfo {volume}
  {46}},\ \bibinfo {pages} {240 } (\bibinfo {year} {1958})}\BibitemShut
  {NoStop}%
\bibitem [{\citenamefont {{Glassgold}}\ \emph {et~al.}(2005)\citenamefont
  {{Glassgold}}, \citenamefont {{Krsti{\'c}}},\ and\ \citenamefont
  {{Schultz}}}]{Glassgold05}%
  \BibitemOpen
  \bibfield  {author} {\bibinfo {author} {\bibfnamefont {A.~E.}\ \bibnamefont
  {{Glassgold}}}, \bibinfo {author} {\bibfnamefont {P.~S.}\ \bibnamefont
  {{Krsti{\'c}}}}, \ and\ \bibinfo {author} {\bibfnamefont {D.~R.}\
  \bibnamefont {{Schultz}}},\ }\href {\doibase 10.1086/427686} {\bibfield
  {journal} {\bibinfo  {journal} {\apj}\ }\textbf {\bibinfo {volume} {621}},\
  \bibinfo {pages} {808} (\bibinfo {year} {2005})}\BibitemShut {NoStop}%
\bibitem [{\citenamefont {Furlanetto}\ and\ \citenamefont
  {Furlanetto}(2007)}]{Furlanetto2007}%
  \BibitemOpen
  \bibfield  {author} {\bibinfo {author} {\bibfnamefont {S.~R.}\ \bibnamefont
  {Furlanetto}}\ and\ \bibinfo {author} {\bibfnamefont {M.~R.}\ \bibnamefont
  {Furlanetto}},\ }\href {\doibase 10.1111/j.1365-2966.2007.11921.x} {\bibfield
   {journal} {\bibinfo  {journal} {Monthly Notices of the Royal Astronomical
  Society}\ }\textbf {\bibinfo {volume} {379}},\ \bibinfo {pages} {130}
  (\bibinfo {year} {2007})}\BibitemShut {NoStop}%
\bibitem [{\citenamefont {Arimondo}\ \emph {et~al.}(1977)\citenamefont
  {Arimondo}, \citenamefont {Inguscio},\ and\ \citenamefont
  {Violino}}]{Arimondo1977}%
  \BibitemOpen
  \bibfield  {author} {\bibinfo {author} {\bibfnamefont {E.}~\bibnamefont
  {Arimondo}}, \bibinfo {author} {\bibfnamefont {M.}~\bibnamefont {Inguscio}},
  \ and\ \bibinfo {author} {\bibfnamefont {P.}~\bibnamefont {Violino}},\ }\href
  {\doibase 10.1103/RevModPhys.49.31} {\bibfield  {journal} {\bibinfo
  {journal} {Rev. Mod. Phys.}\ }\textbf {\bibinfo {volume} {49}},\ \bibinfo
  {pages} {31} (\bibinfo {year} {1977})}\BibitemShut {NoStop}%
\bibitem [{\citenamefont {Ekstrom}\ \emph {et~al.}(1995)\citenamefont
  {Ekstrom}, \citenamefont {Schmiedmayer}, \citenamefont {Chapman},
  \citenamefont {Hammond},\ and\ \citenamefont {Pritchard}}]{Ekstrom1995}%
  \BibitemOpen
  \bibfield  {author} {\bibinfo {author} {\bibfnamefont {C.~R.}\ \bibnamefont
  {Ekstrom}}, \bibinfo {author} {\bibfnamefont {J.}~\bibnamefont
  {Schmiedmayer}}, \bibinfo {author} {\bibfnamefont {M.~S.}\ \bibnamefont
  {Chapman}}, \bibinfo {author} {\bibfnamefont {T.~D.}\ \bibnamefont
  {Hammond}}, \ and\ \bibinfo {author} {\bibfnamefont {D.~E.}\ \bibnamefont
  {Pritchard}},\ }\href {\doibase 10.1103/PhysRevA.51.3883} {\bibfield
  {journal} {\bibinfo  {journal} {Phys. Rev. A}\ }\textbf {\bibinfo {volume}
  {51}},\ \bibinfo {pages} {3883} (\bibinfo {year} {1995})}\BibitemShut
  {NoStop}%
\bibitem [{\citenamefont {Dalgarno}(1961)}]{dal61}%
  \BibitemOpen
  \bibfield  {author} {\bibinfo {author} {\bibfnamefont {A.}~\bibnamefont
  {Dalgarno}},\ }\href {\doibase 10.1098/rspa.1961.0107} {\bibfield  {journal}
  {\bibinfo  {journal} {Proceedings of the Royal Society of London. Series A.
  Mathematical and Physical Sciences}\ }\textbf {\bibinfo {volume} {262}},\
  \bibinfo {pages} {132} (\bibinfo {year} {1961})}\BibitemShut {NoStop}%
\bibitem [{\citenamefont {Landau}\ and\ \citenamefont
  {Lifshitz}(1977)}]{lan77}%
  \BibitemOpen
  \bibfield  {author} {\bibinfo {author} {\bibfnamefont {L.~D.}\ \bibnamefont
  {Landau}}\ and\ \bibinfo {author} {\bibfnamefont {E.~M.}\ \bibnamefont
  {Lifshitz}},\ }\href@noop {} {\emph {\bibinfo {title} {Quantum Mechanics}}}\
  (\bibinfo  {publisher} {Pergamon Press, Oxford},\ \bibinfo {year}
  {1977})\BibitemShut {NoStop}%
\bibitem [{\citenamefont {Hechtfischer}\ \emph {et~al.}(2002)\citenamefont
  {Hechtfischer}, \citenamefont {Williams}, \citenamefont {Lange},
  \citenamefont {Linkemann}, \citenamefont {Schwalm}, \citenamefont {Wester},
  \citenamefont {Wolf},\ and\ \citenamefont {Zajfman}}]{hec02}%
  \BibitemOpen
  \bibfield  {author} {\bibinfo {author} {\bibfnamefont {U.}~\bibnamefont
  {Hechtfischer}}, \bibinfo {author} {\bibfnamefont {C.~J.}\ \bibnamefont
  {Williams}}, \bibinfo {author} {\bibfnamefont {M.}~\bibnamefont {Lange}},
  \bibinfo {author} {\bibfnamefont {J.}~\bibnamefont {Linkemann}}, \bibinfo
  {author} {\bibfnamefont {D.}~\bibnamefont {Schwalm}}, \bibinfo {author}
  {\bibfnamefont {R.}~\bibnamefont {Wester}}, \bibinfo {author} {\bibfnamefont
  {A.}~\bibnamefont {Wolf}}, \ and\ \bibinfo {author} {\bibfnamefont
  {D.}~\bibnamefont {Zajfman}},\ }\href {\doibase 10.1063/1.1513459} {\bibfield
   {journal} {\bibinfo  {journal} {The Journal of Chemical Physics}\ }\textbf
  {\bibinfo {volume} {117}},\ \bibinfo {pages} {8754} (\bibinfo {year}
  {2002})}\BibitemShut {NoStop}%
\end{thebibliography}%

\end{document}